\documentclass[aps,preprint]{revtex4}
\begin{document}
\title{Remarks on Aharonov-Bohm effect and geometric phase}
\author{S.C.  Tiwari}  
\affiliation{Institute  of  Natural  Philosophy \\ 1, Kusum  Kutir, 
Mahamanapuri \\ Varanasi-221005, India} 
\email{vns_sctiwari@yahoo.com}
\begin{abstract}
       Recently the physical  mechanism  for geometric  phase in optics has been
elucidated in terms of the angular momentum holonomy proposed in 1992.  Aharonov
and  Kaufherr  (PRL, 92, 070404,  2004)  revisit the  Aharonov-Bohm  effect, and
propose non-local exchange of a conserved, gauge invariant quantity that changes
the modular momentum of the particle that is responsible for the AB phase shift.
We suggest that the net angular momentum shifts proposed for GP may be analogous
to the shift in the modulus  momentum for the AB effect.  At a single  photon or
electron  level  such  non-trivial,  geometric  effects  seem  to hint  at a new
physics.  
\end{abstract}
\maketitle

(1)   Aharonov   and  Bohm  in  1959  \cite{1}  write  that  in   classical
electrodynamics,  the vector and scalar  potentials  were first  introduced as a
convenient  mathematical  aid for  calculating  fields, and further that, In the
quantum  mechanics,  however, the  canonical  formalism is  necessary,  and as a
result, the potentials  cannot be eliminated  from the basic  equations.  It may
seem  that  even  in  quantum  mechanics,  the  potentials  themselves  have  no
independent   significance.  In  this  paper,  we  shall  show  that  the  above
conclusions are not correct and that a further  interpretation of the potentials
is  needed  in  the  quantum  mechanics.  The  physical  interpretation  of  the
potentials  has been  debated  since  their  introduction,  and as early as 1861
Riemann  interpreted them  representing  the density and velocity of aether.  In
the monograph \cite{2}, Chapter 6 we have reviewed  modifications/extensions 
of the Maxwell-Lorentz  theory, and in a recent paper the physical attributes
of photon are discussed in terms of the  potentials  \cite{3}.  Note that 
classical or quantum mechanical  description of a physical  phenomenon is just
a description, and if, there  are   unexplained   or  puzzling   aspects  of 
physical   reality   mere correspondence between the descriptions or setting
the limits of validity of the theories  would not be enough:  search for a new
theory becomes  imperative.  It is this belief that has led us to  re-examine 
the concepts of space and time in Newtonian  and  relativistic  dynamics,  and
the  structure  of  Maxwell-Lorentz electrodynamics for single electron and
photon \cite{2}.  In contrast to Aharonov and Bohm  proposition  that
potentials have  significance  in QM, we argue that at a basic level the 
potentials  describe a single  photon, and the  electromagnetic field is a
macroscopic  quantity  representing  the photon fluid.  Here magnetic field 
signifies  the angular  momentum of this  fluid.  Note that the  dynamics
based on the rate of change of momentum  cannot  describe the effect of
constant potentials  for a force  derivable from a potential (as its 
gradient).  We call the force-free (or torque-free) dynamics as pre-dynamics,
and suggest that under certain  conditions  these  pre-dynamical  effects  of 
unobservable  potentials manifest as changes in the level of momentum  (or
angular  momentum);  AB effect and GP have origin in such  shifts.  The 
physical  meaning of angular  momentum holonomy and possible experimental
tests are discussed in \cite{4}.  

(2) Recent paper \cite{5}  discussing  the AB effect  underlines  the fact 
that in spite of  several interpretations  of this effect, there do exist 
intriguing  features.  Aharonov and Kaufherr \cite{5} compare the classical 
Poisson  bracket  formalism  and quantum Heisenberg  equation of motion for a
charged  particle  motion in the field-free region, and find that there exists
a quantity that changes in the later case but does not change in the former. 
Authors argue that in quantum theory this change occurs in the modular 
momentum, and the exchange occurs at a definite time.  An experimental  test
is  proposed to test this idea making use of an  ensemble  of charged 
particles  and the effect of gauge  invariant  exchange on the velocity
distribution.  In their analysis \cite{5}, the displacement  operator is
compared for classical and quantum  descriptions.  The displacement operator
is a function of modular momentum.  For the electromagnetic waves, the
classical treatment of the interference  depends on the relative  phase, while
for the quantized  field the operator  for {\bf E} $\times$ {\bf B} is used to
define the  displacement  operator.  It is noted, that thus, for a single
photon in a superposition of two wave packets  displaced by L, this operator
depends on the relative phase in exactly the same way as the modular  momentum 
considered  above does From this remark it seems the  modular momentum is
important for single photon, not for classical  light beams.  On the other
hand, typical  geometric  phases occur for classical  light, and there are
claims   that  they  have  been   measured   for   single   photons.  A 
natural generalization  of the  analysis  given in \cite{5}  would be to 
construct  rotation operators and analogous to the modular momentum define
modular angular momentum. If the modular  effects are  important  only for
single  photons,  then it would imply that GP cannot be related with angular
momentum holonomy as proposed by us \cite{4}.  However, from the  discussion on
the AB effect, we prefer to interpret the modular angular momentum as
representing the angular momentum shifts responsible for GP.  We do not like
to treat physical effects as  description-specific:  for example,  though
Hannay angles \cite{6} arise in one of the  formalisms  of classical mechanics,
namely the action-angle  variables, the observed effect is a property of  the 
physical  system/process,  not an  artifact  of  the  description.  

(3) Preceding discussion leads to two fundamental questions:  what is the
meaning of single photon?  What is phase?  In ref.3 we have critically
reviewed the concept of photon; we add to that logically  sound arguments for
anti-photon by Lamb \cite{7} questioning  the  scientific  justification  for
the  photon  concept  in  laser physics.  In our review \cite{3} we have 
pointed  out that the  concept of photon in quantum optics and quantum 
electrodynamics  remains a  calculational  tool, and therefore its physical
reality is at best undecidable.  If we demand elimination of  unsound 
mathematical   procedures  (e.g.  subtraction  of  infinities)  and
counter-intuitiveness  of quantum theory from quantum  electrodynamics, then
not only the  classical  theory may have to be revised,  photon and 
electromagnetic potentials  may turn out to be  fundamental.  In the last
section of \cite{3} we have outlined a tentative  experimental  schematic to
test the photon  model in which the magnetic flux quantum is re-interpreted as
photon spin and electronic charge as fractional spin.  The  polarization 
entangled photon pairs are sent into the two arms of the  interferometer; in
one of the arms, the optical fiber is passed through a  superconducting  ring
placed  between the analyzer and the  detector. The  correlations  are
recorded, and compared for the two cases when the ring is in the  normal 
state and when it is cooled to the  superconducting  state.  The difference 
between the two should  show up as photon  trapping,  and be related with the 
number  of flux  quanta.  Re-interpreting  classical  light as  photon fluid,
we have also suggested application of static magnetic field to change the
orbital  angular  momentum of light beams, see \cite{3}.  As regards to the 
physical significance  of the phase, we refer to \cite{4} and propose to
discuss  this problem in connection with the rotational frequency shifts in
the sequel.

\end{document}